\documentclass[useAMS,usenatbib,usegraphicx]{mn2e}
\usepackage{amsmath,amsfonts,amssymb,url,pslatex,color,float}
\def\idm#1{{\mbox{\scriptsize #1}}}

\def\kepler{{\em Kepler}}
\def\Y{\langle Y \rangle}

\def\mean#1{\langle{}#1{}\rangle}
\def\tv#1{{\pmb #1}}

\newcommand{\au}{\mbox{au}} % IAU convention
\newcommand{\msun}{\mbox{m}_{\odot}}

\newcommand{\mE}{\mbox{m}_{\oplus}}
\newcommand{\mJ}{\mbox{m}_{\idm{Jup}}}
\newcommand{\RE}{\mbox{R}_{\oplus}}
\newcommand{\Mmean}{\mathcal{M}}

\newcommand{\tepoch}{t_0}

\newcommand{\mb}{m_{\idm{1}}}
\newcommand{\mc}{m_{\idm{2}}}
\newcommand{\md}{m_{\idm{3}}}

\newcommand{\Pd}{P_{\idm{3}}}

\newcommand{\yr}{\mbox{yr}}
\newcommand{\ab}{a_{\idm{1}}}
\newcommand{\ac}{a_{\idm{2}}}

\newcommand{\eb}{e_{\idm{1}}}

\def\kepler{{\sc Kepler}}
\def\deg{{\rm o}}
\def\idm#1{{\mbox{\scriptsize #1}}}
\newcommand\Chi{{\chi^2_\nu}}

\title[The Laplace resonance in the Kepler-60 system]
{The Laplace resonance in the Kepler-60 planetary system}
\author[K.~Go\'zdziewski, C.~Migaszewski, F.~Panichi \& E.~Szuszkiewicz]%
{K. Go\'zdziewski$^{1}$, C. Migaszewski$^{1,2}$, 
F. Panichi$^{2}$, and E. Szuszkiewicz$^{2}$ \\
$^{1}$Centre for Astronomy, Faculty of Physics, Astronomy and Informatics,
Nicolaus Copernicus University, Grudziadzka 5, 87-100 Toru\'n, Poland \\
$^{2}$Institute of Physics and CASA*, Faculty of Mathematics and Physics,
University of Szczecin,  Wielkopolska 15, 70-451 Szczecin, Poland}
\begin{document}
%_______________________________________________________________________________
%
\maketitle
\begin{abstract}
We investigate the dynamical stability of the Kepler-60 planetary system with
three super-Earths. We determine their orbital elements and masses by
Transit Timing Variation (TTV) data spanning quarters {Q1-Q16} of the
\kepler{} mission. The system is dynamically active but the TTV data constrain
masses to $\sim4~\mE$ and orbits in safely wide stable zones. The observations
prefer two types of solutions. The true three-body Laplace MMR exhibits the
critical angle librating around $\simeq 45^{\circ}$ and aligned apsides of the
inner and outer pair of planets. In the Laplace MMR formed through a chain of
two-planet 5:4 and 4:3 MMRs, all critical angles librate with small amplitudes
$\sim30^{\circ}$ and apsidal lines in planet's pairs are anti-aligned.
The system is simultaneously locked in a three-body MMR with librations
amplitude $\simeq 10^{\deg}$. The true Laplace MMR can evolve towards a chain of
two-body MMRs in the presence of planetary migration. Therefore {the
three-body MMR} formed in this way seems to be more likely state of the system.
However, the true three-body MMR cannot be disregarded {\em a priori} and it
remains a puzzling configuration that may challenge the planet formation theory.
\end{abstract}
%_______________________________________________________________________________
%
\date{Accepted .... Received ...; in original form ...}
\pagerange{\pageref{firstpage}--\pageref{lastpage}} \pubyear{2012}
\maketitle
\label{firstpage}
%_______________________________________________________________________________
%
\section{Introduction}
%_______________________________________________________________________________
%
The \kepler{} mission has lead to the discovery of a~few hundred multiple
planetary systems with super-Earth planets. Such systems may be involved in
low-order mean motion resonances (MMRs) or close to the MMRs
\citep[e.g.][]{Fabrycky2013}. The Transit Timing Variation \citep{Agol2005} or
the photodynamical $N$-body method \citep{Carter2011} are common approaches
making it possible to determine planetary masses and orbital architectures of
their parent systems. A large sample of \kepler{} multiple systems has been
analyzed in \citep{Rowe2015,Mullally2015}. They provided a rich catalogue of
basic photometric models {of} such systems, including TTV measurements for
a number of them.

In this sample, the Kepler-60 system of three super-Earth planets
\citep{Steffen2012} with yet unconstrained masses reveals the mean orbital
period ratios $\sim1.250$ and $\sim1.334$ very close to relatively prime
integers. This is suggestive for a chain of~two-body MMRs. One pair of planets
is near the 5:4~MMR and the second one is near the 4:3 MMR. Moreover, the
mean-motions of Kepler-60b ($n_{1}$), Kepler-60c ($n_{2}$) and Kepler-60d
($n_{3}$) satisfy 
$
n_{1} - 2 n_{2} + n_{3} \approx {-0.002^{\circ}{\rm d}^{-1}}.
$
This relation may be understood as the generalized Laplace resonance
\citep{Papaloizou2014}.

Our preliminary dynamical analysis has revealed that the mutual interactions
between the Kepler-60 planets are strong and the dynamical constraints may be
decisive for maintaining the long-term stability of this system. Therefore,
besides a characterization of the putative resonance, and determining the
masses, we aim to perform a comprehensive dynamical study of the Kepler-60 on
the basis of {the TTV measurements} in \citep{Rowe2015}. 

%
%_______________________________________________________________________________
%
\section{Three-body mean motion resonance}
%_______________________________________________________________________________
%
Three-body MMR's may be considered as the most important case of multiple MMRs,
after two-body MMRs, which may actively shape short-term and long-term
dynamics of multi-planet configurations. The overlap of two-body and three-body
MMRs has been identified as a source of deterministic chaos in the Outer Solar
system and {in the asteroid belt
\citep[e.g.][]{Murray1999,Guzzo2006,Smirnov2013}}. The best-studied and known
example of the three-body MMR is the Laplace resonance of the Galilean moons.
The mean-motions of Io ($n_\idm{Io}$), Europa ($n_\idm{E}$) and Ganymede
($n_\idm{G}$) satisfy 
$
n_\idm{Io} - 2n_\idm{E} = n_\idm{E} - 2n_\idm{G} \approx 0.74^{\circ}{\rm d}^{-1}
$
that implies
$n_\idm{Io} -3 n_\idm{E} + 2 n_\idm{G}~\simeq 0^{\circ}{\mbox d}^{-1}$  
\citep[e.g.][]{Sinclair1975}.

Following \cite{Papaloizou2014}, we consider a subset of {five} types of the
Laplace MMRs. In a coplanar system, a combination of the first-order MMRs'
librating--nonlibrating critical angles $\phi_{1,j} = p \lambda_{1} -(p+1)
\lambda_{2} + \varpi_{j}$, $\phi_{2,j} = q \lambda_{2} -(q+1) \lambda_{3} +
\varpi_{j+1}$, for relatively prime integers $p, q>0$, where $j=1,2$,
$\lambda_{i}$, $\varpi_{i}$ are the mean longitudes and pericenter longitudes
$(i=1,2,3)$ for subsequent planets, results in a chain of two-body MMRs of the
first order. Librations of the critical angles in this chain naturally imply
librations of 
$
\phi_L =p \lambda_{1} - (p+q+1) \lambda_{2} + (q+1) \lambda_{3}
$
(type I). A scenario when only $\phi_L$ librates but all two-body MMR critical
angles circulate will be called the ``true'' or ``pure'' three-body MMR (type
II). We do not consider here two more types related to three or two out of four
two-planet MMR critical angles librating.

The first extrasolar system exhibiting the three-body Laplace MMR has been
discovered around Gliese 876 \citep{Marcy2001,Rivera2010}. In this case two-body
MMR critical angles of the first pair of planets librate around 0$^{\circ}$, one
ciritcal angle librates around 0$^{\circ}$ for the outer pair and $\phi_L$
librates around 0$^{\circ}$ with amplitude $\sim40^{\circ}$\citep{Marti2013}.
Another intriguing example of {\em two instances} of the Laplace resonance in
one planetary system is likely the HR~8799 system of four massive $\sim10~\mJ$
planets in wide orbits $\sim100$~au \citep{Marois2010,Gozdziewski2014}. Given a
putative Laplace resonance in the Kepler-60 system, this MMR may form in
different environments, spanning wide mass-ranges{, from tiny moons of
Pluto \citep{Showalter2015}, the Mercury-like satellites of Jupiter,
super-Earths and Jovian planets, to (and perhaps beyond) the brown dwarf mass
limit.}
  
%_______________________________________________________________________________
%
\section{The TTV data model and optimization}
%_______________________________________________________________________________
%
The TTV dataset ${\cal D}$ for Kepler-60 consists of 419 measurements spanning
quarters Q1-Q16, with a reported mean uncertainty $\mean{\sigma} \sim
0.024$~days \citep{Rowe2015}. Due to this large scatter which compares to the
magnitude of the TTV data, we introduce a simplified orbital architecture. We
assume that the system is coplanar or close to coplanar and then the
inclinations $I_i=90^{\circ}$ and nodal longitudes $\Omega_i=0^{\circ}$,
$i=1,2,3$. Orbits in compact \kepler{} systems are usually quasi-circular, hence
to get rid of weakly constrained longitudes of pericenter $\varpi_i$ when
eccentricities $e_i\sim0$, we introduce osculating, astrocentric elements
$\left\{ P_i, x_i, y_i, T_i \right\}$ instead of $\left\{a_i, e_i, \omega_i,
\Mmean_i \right\}$:
\[
P_i = 2\,\pi \sqrt{\frac{a_i^3}{k^2\,(m_0 + m_i)}}, \quad
T_i = \tepoch + 
\frac{P_i}{2\pi} \left( \Mmean_i^{(\idm{t})} - \Mmean_i \right),
\]
and $x_i = e_i \, \cos\varpi_i$, $y_i = e_i \, \sin\varpi_i$
where $k$ is the Gauss constant, $\Mmean_i^{(\idm{t})}$ is the mean
anomaly at the epoch of the first transit $T_i$, and ${\cal M}_i$, $P_i$, $a_i$
are for the mean anomaly, the orbital period and semi-major-axis, at the
osculating epoch $\tepoch$ for each planet, respectively. We computed the
transits {moments with a code by \cite{Deck2014}}
and with our own code for an independent check.

{The TTV technique is plagued by non-unique and/or unconstrained solutions.
Therefore, we applied two different methods of quasi-global optimization in
parallel, to explore the space of free parameters $\tv{\theta}$ in the model
including geometric elements $(P_i, T_i, x_i, y_i)$ and masses $m_i$, for
$i=1,2,3$ marking subsequent planets, as well as the measurements uncertainty
correction term $\sigma_f$ (see below). }

{
The Markov Chain Monte Carlo (MCMC) technique is widely used by the photometric
community to determine the posterior probability distribution ${\cal
P}(\tv{\theta}|{\cal D})$ of model parameters $\tv{\theta}$, given the data set
${\cal D}$: 
$
  {\cal P}(\tv{\theta}|{\cal D}) 
  \propto {\cal P}(\tv{\theta}) {\cal P}({\cal D}|\tv{\theta}),
$
where ${\cal P}(\tv{\theta})$ is the prior, and the sampling data distribution
${\cal P}({\cal D}|\tv{\theta}) \equiv \log{\cal L}(\tv{\theta},{\cal D})$. We
choose all priors as flat (or uniform improper) by placing limits on model
parameters, i.e., $P_{i}>0$~d, $T_{i}>0$~d, $x_i,y_i \in (-0.5,0.5)$, $m_i \in
[0.1,30]~\mE$, $\sigma_f>0$~d ($i=1,2,3$). }

Since our preliminary fits revealed $\Chi \sim1.7$ and a large scatter of
residuals suggestive for underestimated uncertainties, we
optimized the maximum likelihood function ${\cal L}$:
\begin{equation}
 \log {\cal L} =  
-\frac{1}{2} \sum_{i,t}
\frac{{\mbox{(O-C)}}_{i,t}^2}{\sigma_{i,t}^2}
- \frac{1}{2}\sum_{i,t} \log {\sigma_{i,t}^2} 
- \frac{1}{2} N \log{2\pi},
\label{eq:Lfun}
\end{equation}
where $(\mbox{O-C})_{i,t}$ is the (O-C) deviation of the observed $t$-th transit
moment of an $i$-th planet from its $N$-body ephemeris, and the TTV uncertainty
$\sigma_{i,t}^2 \rightarrow \sigma_{i,t}^2+\sigma_f^2$ with $\sigma_f$ parameter
scaling the raw uncertainty $\sigma_{i,t}$ and $N$ is the total number of TTV
observations. We assume that the uncertainties are Gaussian and independent.
Indeed, {\em a posteriori} Lomb-Scargle periodograms of the residuals of
best-fitting models did not show significant frequencies.

Because the values of ${\cal L}$ are non-intuitive for comparing solutions, a
rescaled value 
$
 \log L = \log 0.2420 - \log {\cal L}/N
$
(in days) is more suitable to express the quality of fits. Since the
best-fitting models should provide $\chi^2/N \sim1$, then $L \sim
\mean{\sigma}$ measures a scatter of measurements around the best-fitting
models, similar with an r.m.s --- smaller $L$ means better fit
\citep[e.g.,][]{Baluev2009}.

{To perform the MCMC analysis we used the affine-invariant ensemble sampler
\citep{Goodman2010,Foreman2014}. As a different method,} we applied a few kinds
of genetic and evolutionary algorithms \citep[GEA from
hereafter,][]{Charbonneau1995,Izzo2010} to independently maximize the $\log
{\cal L}$ function. We set the same parameter bounds in the GEA runs as in the
MCMC experiments.

%_______________________________________________________________________________
%
\subsection{The best-fitting configurations}
%_______________________________________________________________________________
%
We first performed an extensive search with the GEA, collecting $\sim10^6$
solutions in each multi-CPU run. We found that the best-fitting models with $L
\simeq 0.029$~d have well determined minima for the orbital periods $P_i$ and
{transit epochs $T_i$}, similar with the error term $\sigma_f \sim
0.02$~days. As expected, eccentricities are not constrained and we found equally
good models in whole $(x_i,y_i)$-planes.

The results are consistent with the MCMC experiments illustrated in {a
figure in supplementary material}. It shows one-- and two--dimensional projections
of the posterior probability distribution. 
The posterior is multi-modal and
complex with a dominant peak localized roughly around masses of $\sim4~\mE$ and
a weaker peak around $\sim10~\mE$ for all planets. There are also two relatively
weak, quasi-symmetric peaks for all $x_i\sim\pm0.25$ and $y_i\sim0.25$. Strong
linear correlation between all pairs of $(x_i,x_j)$ and $(y_i,y_j)$, $i\neq j$
is present. Given close commensurabilities of the orbital periods, it is an
additional indication of the MMRs in the system. The linear correlations mean a
tight alignment of apsidal lines which is a typical feature of the low-order
MMRs. 

\begin{figure}
\centerline{
\hbox{
\vbox{
\includegraphics[width=0.48\textwidth]{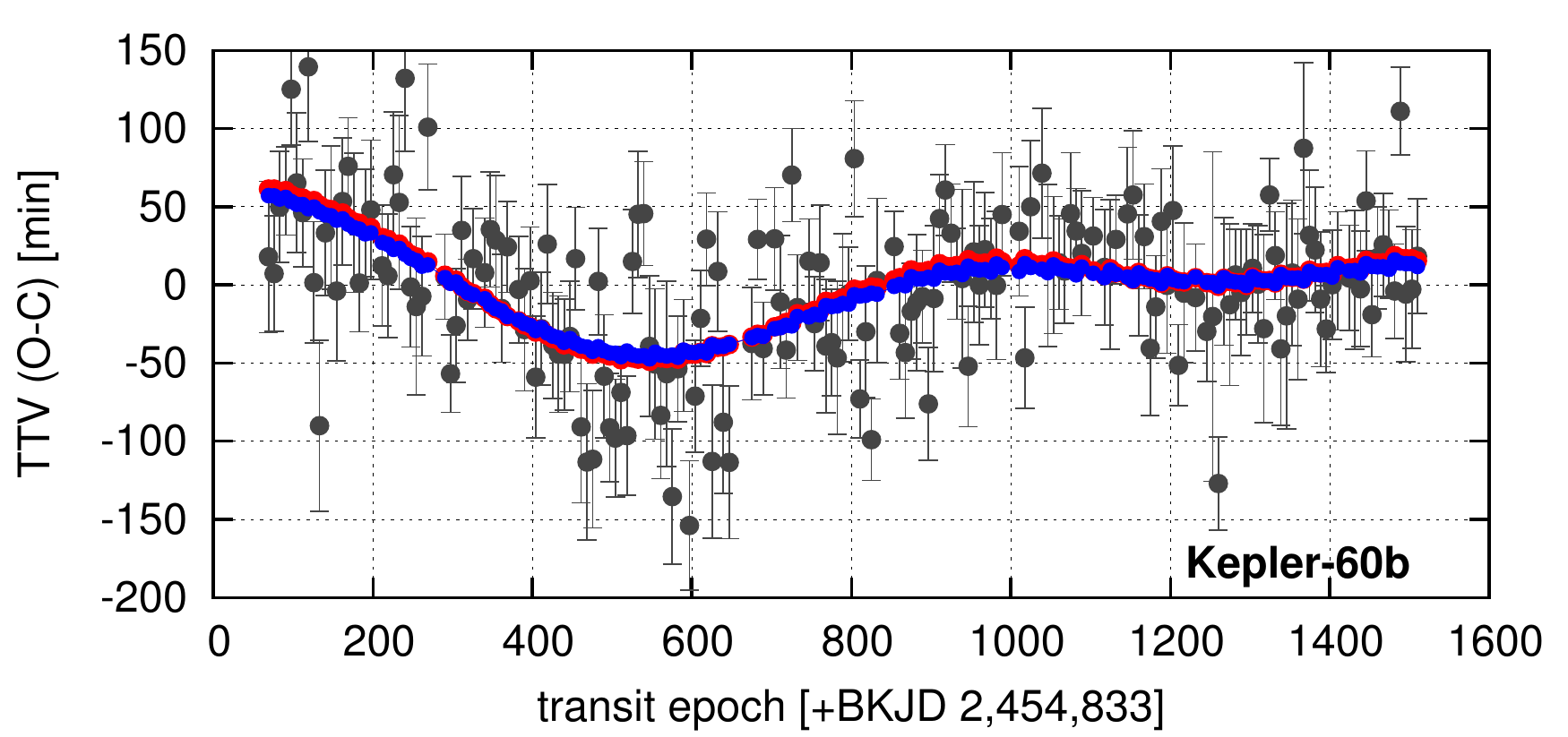}
\includegraphics[width=0.48\textwidth]{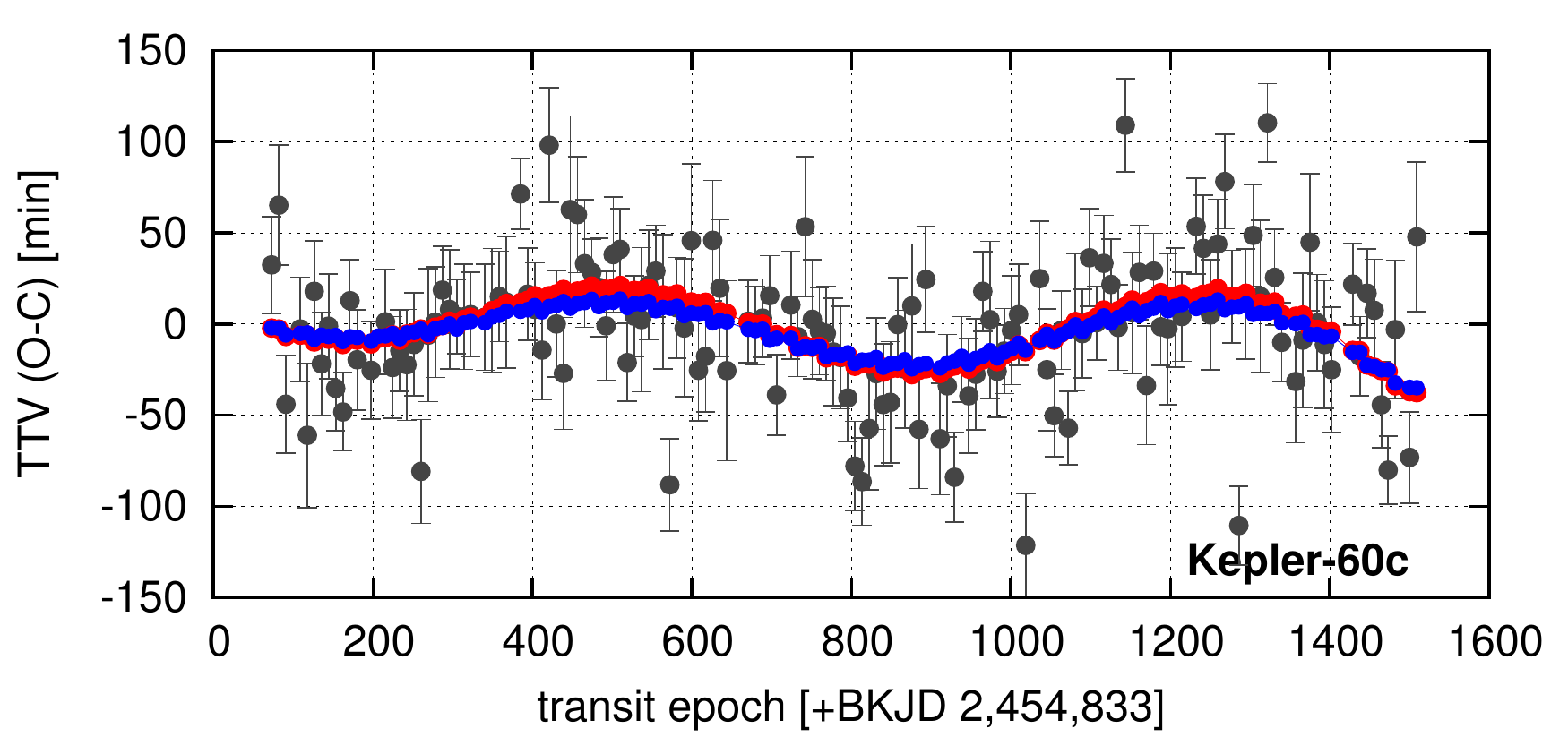}
\includegraphics[width=0.48\textwidth]{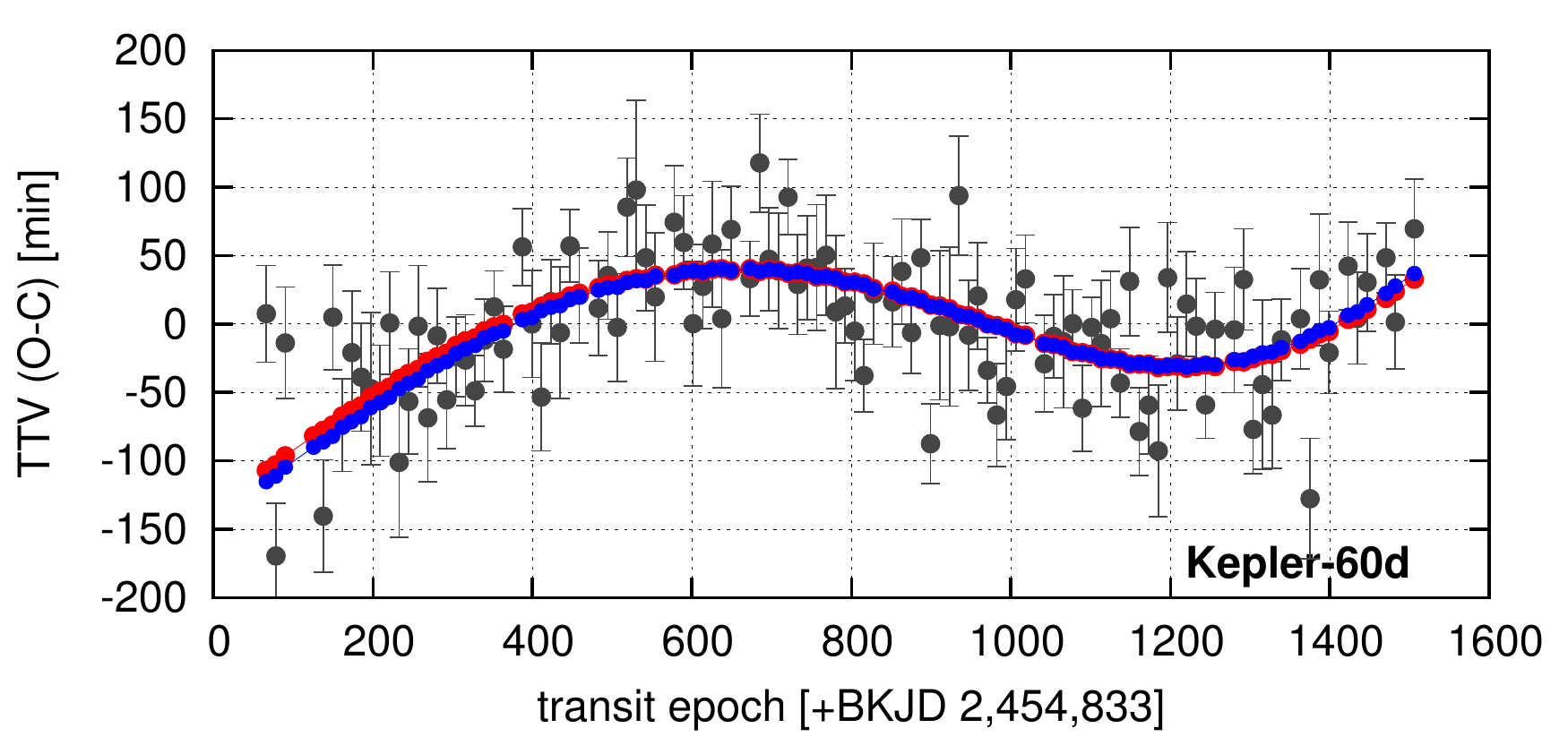}
}
}
}
\caption{
Synthetic curves of best-fitting low-eccentricity models (Tab.~\ref{tab:tab1}):
Fit~I (red circles) and Fit~II (blue circles) over-plotted on the TTV data.
}
\label{fig:fig1}
\end{figure}

\begin{figure}
\centerline{
\hbox{
\vbox{
\includegraphics[width=0.47\textwidth]{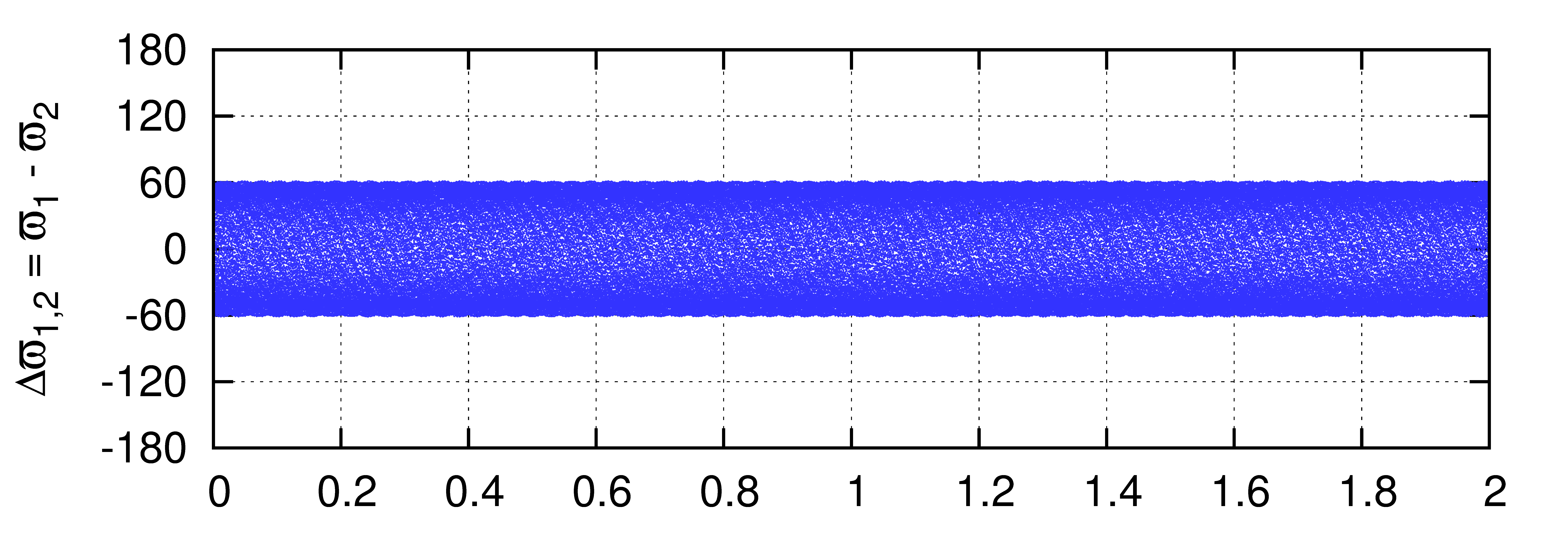}
\includegraphics[width=0.47\textwidth]{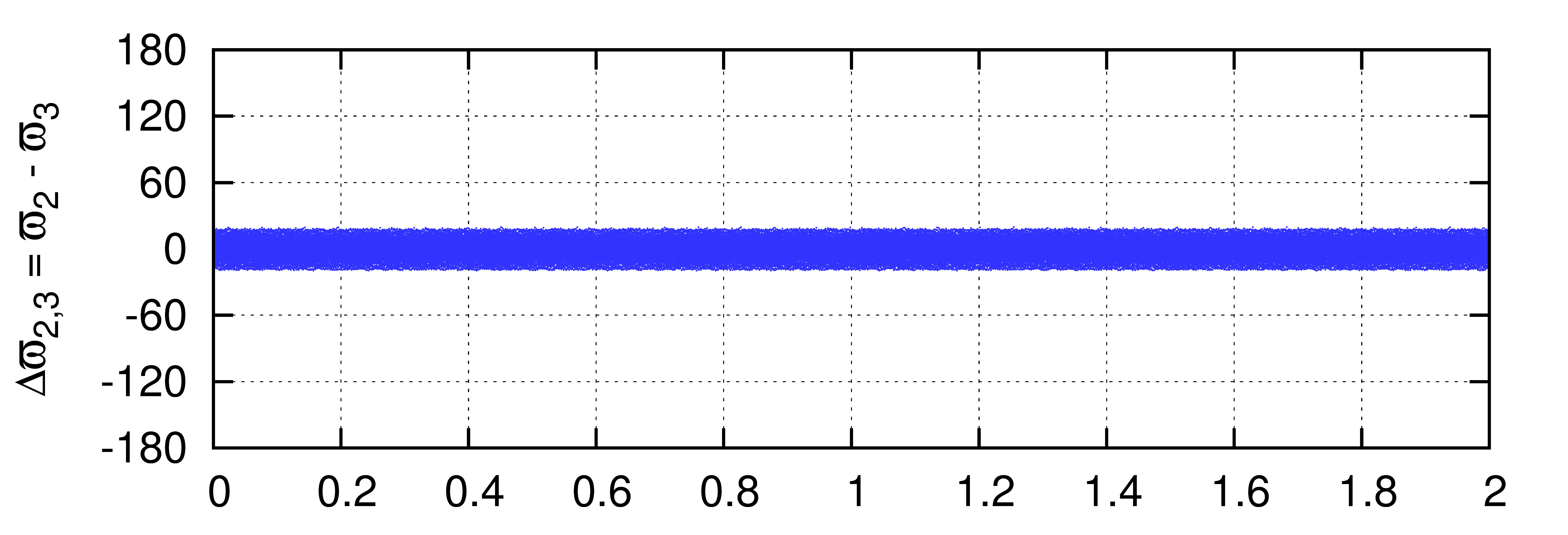}
\includegraphics[width=0.47\textwidth]{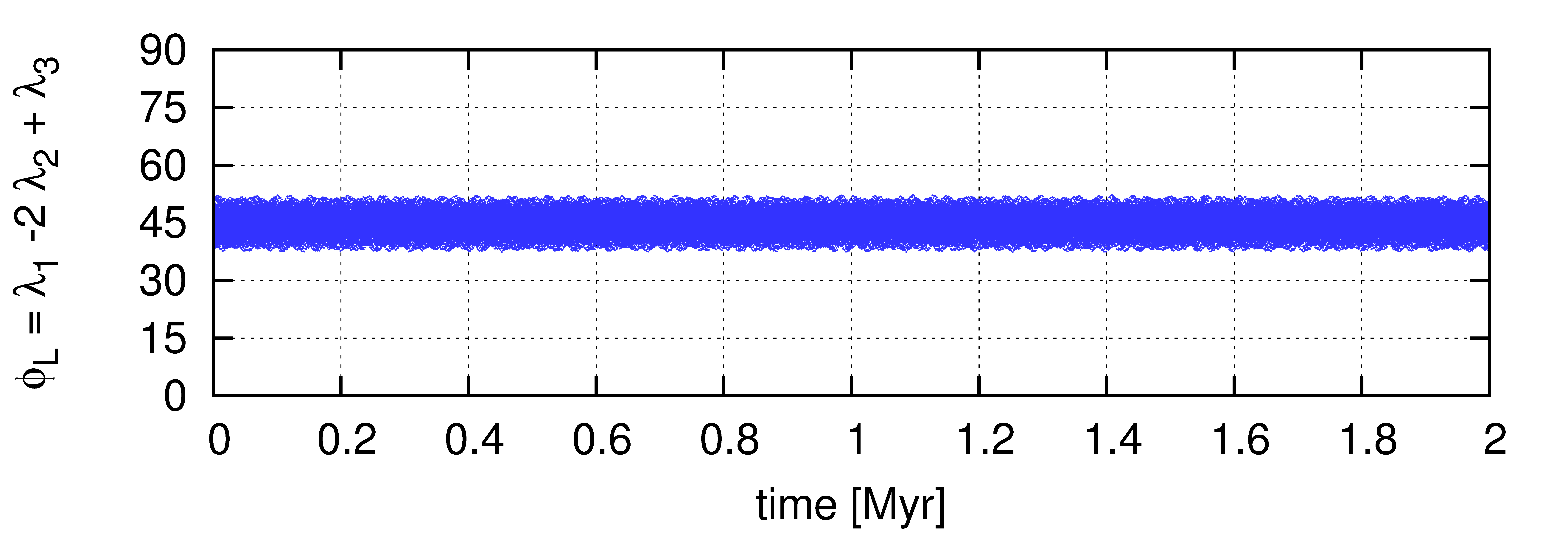}
}
}
}
\caption{
{Evolution of the Laplace MMR critical angle $\phi_L$ and apsidal angles
$\Delta\varpi_{i,j}$ for Fit~I (Tab.~1). All two-body MMR critical angles rotate
(not shown).}
}
\label{fig:fig2}
\end{figure}

\begin{figure}
\centerline{
\hbox{
\vbox{
\includegraphics[width=0.47\textwidth]{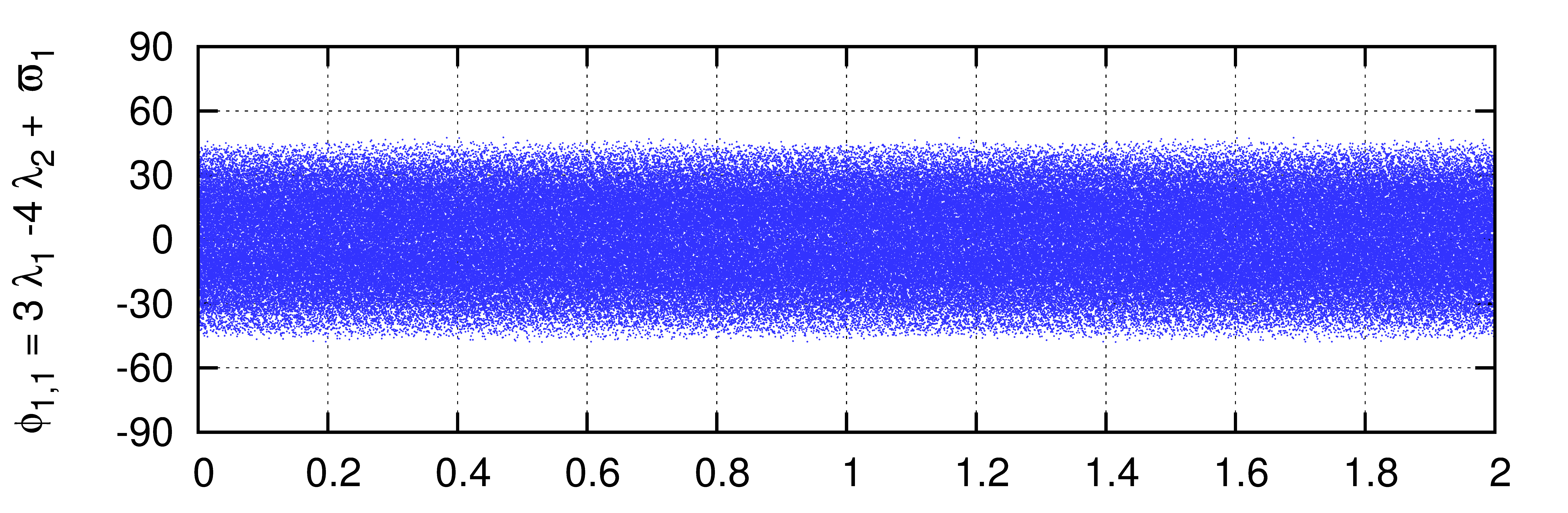}
\includegraphics[width=0.47\textwidth]{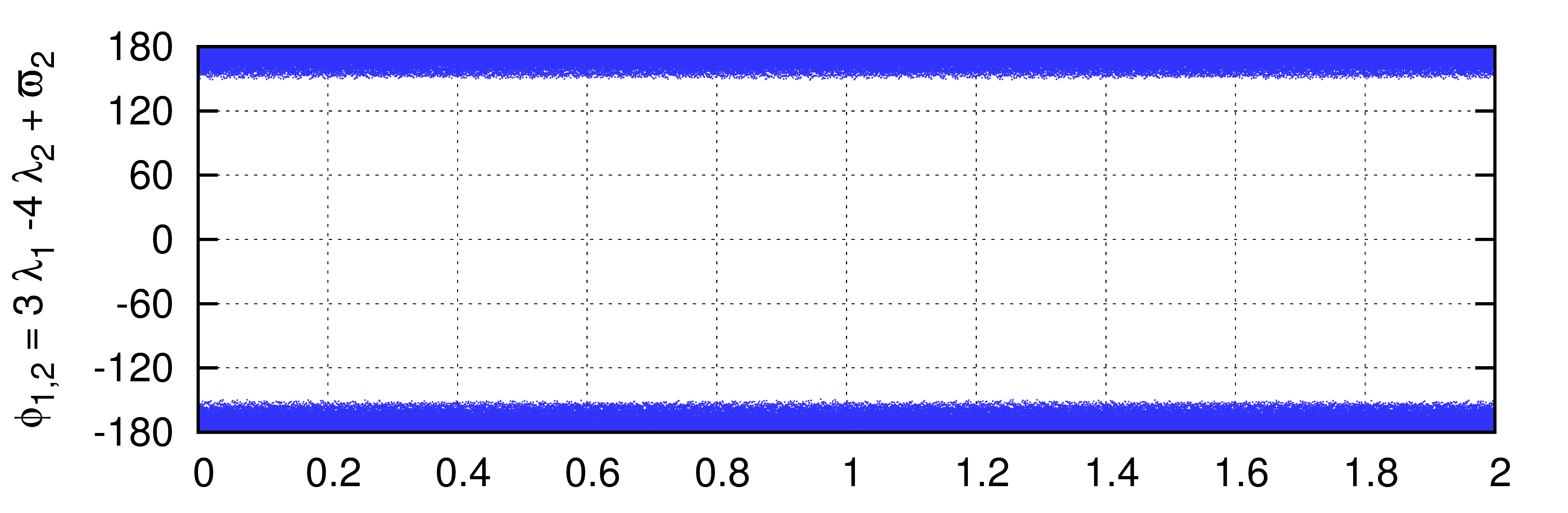}
\includegraphics[width=0.47\textwidth]{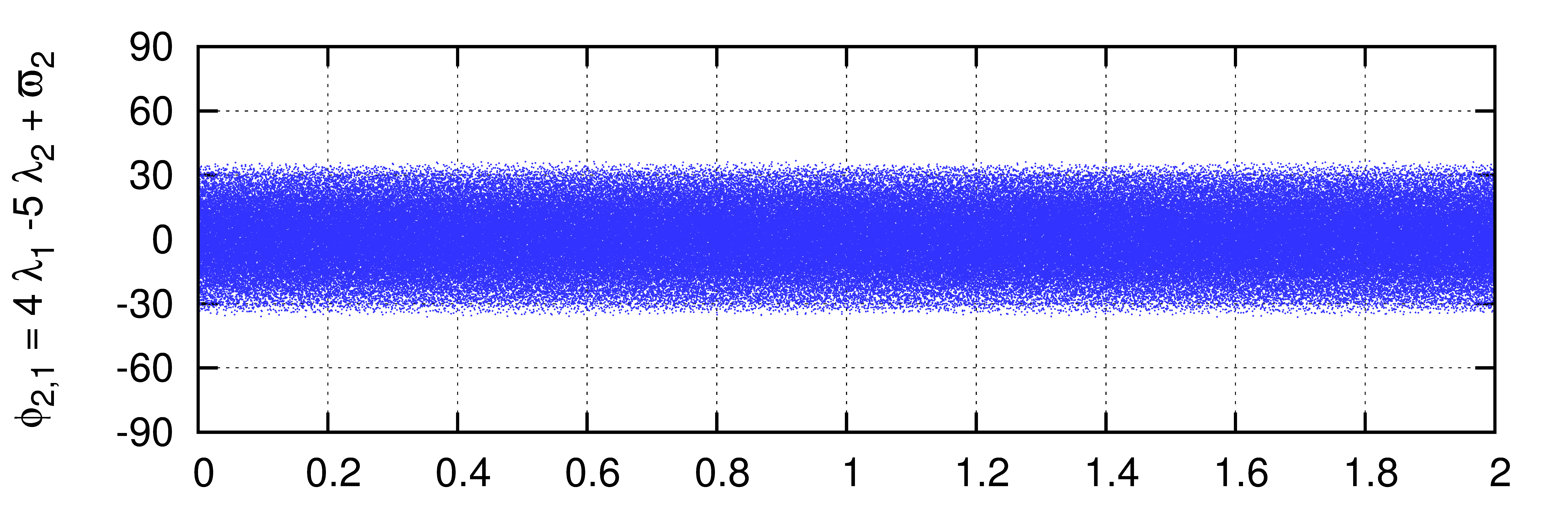}
\includegraphics[width=0.47\textwidth]{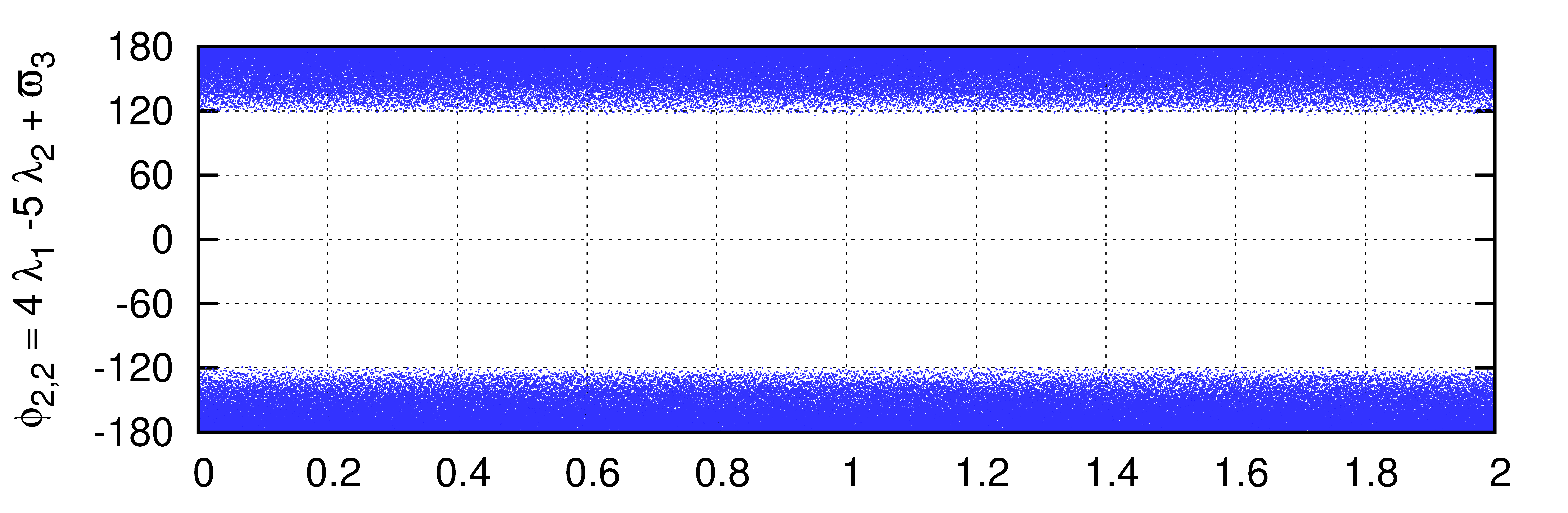}
\includegraphics[width=0.47\textwidth]{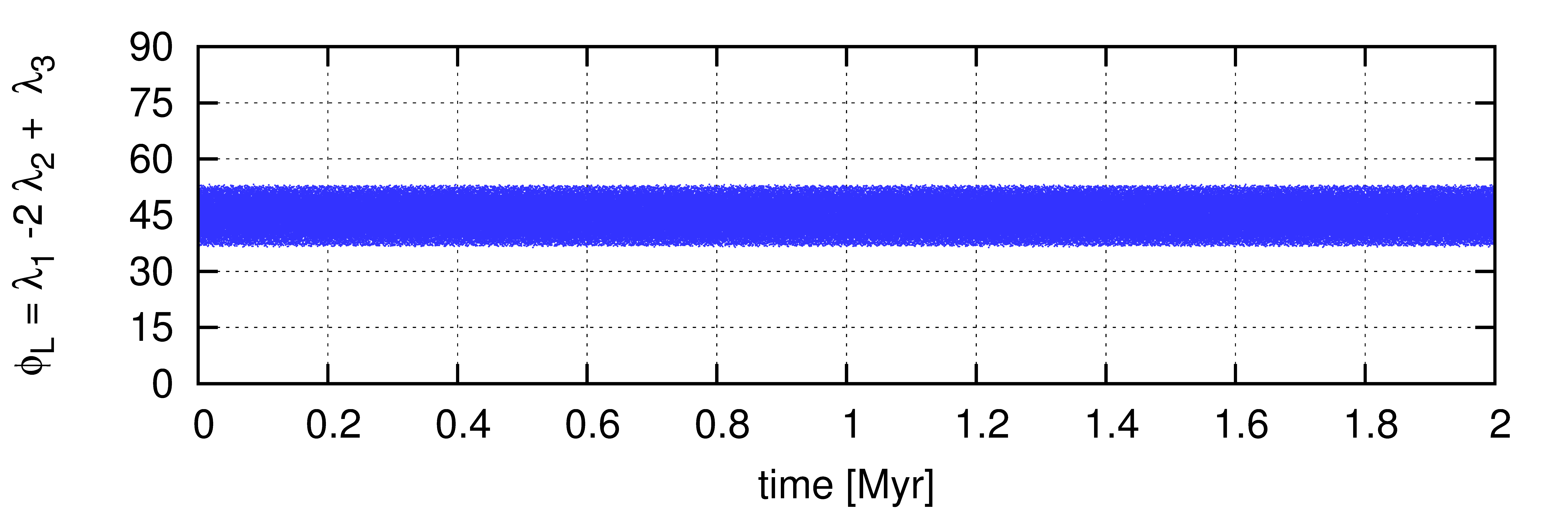}
}
}
}
\caption{
{
Evolution of the critical angles for Fit~II (Tab.~1) corresponding to a chain of
two-body MMRs. Apsidal angles $\Delta\varpi_{1,2}$ and $\Delta\varpi_{2,3}$
librate around $180^{\circ}$ with semi-amplitudes $\sim30^{\circ}$ (not shown
here).
}
}
\label{fig:fig3}
\end{figure}

We searched for other signatures of the MMRs by computing amplitudes of the
critical angles of the two-body and three-body MMRs in the GEA runs, by
integrating numerically the $N$-body equations of motion for all models with
$L<0.050$~d. We found that basically all the solutions exhibit librations of
$\phi_L = \lambda_{1} -2 \lambda_2 + \lambda_3$ around $\simeq 45^{\circ}$ with
amplitudes as small as $10^{\circ}$. Most of these models do not show librations
of any of the two-body MMRs, hence they represent the true three-body MMRs. For
small eccentricities we found similar small $L\sim0.029$~d, low-amplitude
$\phi_L$ solutions with all two-body MMR critical angles librating. That means a
chain of two-body MMRs. Elements of two representative low-eccentricity models
with small $L\simeq 0.029$~d are given in Table~\ref{tab:tab1} and illustrated
in Fig.~\ref{fig:fig1}. These qualitatively different modes of the Laplace MMR
(marked in red and blue, respectively) can be hardly distinguished by the fit
quality. {We note that within uncertainties these solutions have the same
$P_i$, $T_i$ and $m_i$. Regular evolutions of the critical angles as well as
apsidal angles $\Delta\varpi_{i,j} = \varpi_i-\varpi_j$ for 2~Myrs are shown in
Figs.~\ref{fig:fig2} and~\ref{fig:fig3}, respectively.} 

\begin{table}
\caption{
Orbital parameters of representative models of the Kepler-60 system for the true
three-body Laplace resonance (Fit I) and for a chain of two-body MMRs (Fit II).
The osculating epoch is BKJD+65.0. Both systems are coplanar with the
inclination $i=90^{\circ}$ and nodal longitudes $\Omega=0^{\circ}$. Mass of the
star is $1.105\,\msun$ \citep{Rowe2015}. Elements $x_i,y_i$ and $e_i,\varpi_i$
are weakly {limited with the TTV, but are constrained by the dynamical
stability}.
}
\label{tab:tab1}
\begin{tabular}{l r r r}
\hline
planet           & Kepler-60\,b & Kepler-60\,c & Kepler-60\,d \\
\hline\hline
\multicolumn{4}{c}{Fit~I (three-body true Laplace MMR)} \\
\hline
$m_p\,[\mE]$     & 4.0$\pm$0.7  &   5.0$\pm$1.0   & 3.6$\pm$1.1 \\
$P\,$[d]         &  7.1320$\pm$0.0005 & 8.9179$\pm$0.0006 & 11.903$\pm$0.001 \\
$e \cos\varpi$   &  0.0152    &   0.0512  & 0.0077 \\
$e \sin\varpi$   &  -0.0354   &    -0.0244  & -0.0204 \\
$T\,$[d]         & 69.032$\pm$0.009 & 73.075$\pm$0.009  &  66.267$\pm$0.012  \\
$a\,[\au]$       &  0.07497   &   0.08701  & 0.10548 \\
$e$              &   0.0386   &   0.0567  & 0.0218  \\
$\omega\,$[deg]  &  -66.74    &  -25.47   & -69.44 \\
${\cal M}\,$[deg]&  134.92    &   -24.75  & 301.99  \\
$\sigma_f$\,[d]  & \multicolumn{3}{c}{0.018 $\pm$ 0.002} \\
$L$\,[d]         & \multicolumn{3}{c}{0.029} \\
\hline
\multicolumn{4}{c}{Fit~II (two-body MMR chain with Laplace MMR)} \\
\hline
$m_p\,[\mE]$     &  4.1$\pm$0.7   &   4.8$\pm$1.0  &  3.8$\pm$1.1 \\
$P\,$[d]         & 7.1320 $\pm$0.0005 & 8.9177$\pm$0.0006 & 11.903$\pm$0.001 \\
$e \cos\varpi$   & -0.0108    &  0.0282   &  -0.0129 \\
$e \sin\varpi$   & 0.0008     &  0.0073   & 0.0080 \\
$T\,$[d]         &  69.035$\pm$0.009 & 73.075$\pm$ 0.009 & 66.273$\pm$0.012 \\
$a\,[\au]$       &  0.07497    &   0.08701  & 0.10548 \\
$e$              &  0.0108    &   0.0291  & 0.0151 \\
$\omega\,$[deg]       &  175.56    &   14.61  &  148.19 \\
${\cal M}\,$[deg]    &    -110.47  &  -67.35   &  81.84 \\
$\sigma_f$\,[d]  & \multicolumn{3}{c}{0.018 $\pm$ 0.002} \\
$L$\,[d]         & \multicolumn{3}{c}{0.029} \\
\hline
\end{tabular}
\end{table}

%_______________________________________________________________________________
%
\subsection{The long-term stability of resonant configurations}
%_______________________________________________________________________________
%
To investigate the dynamical stability of the system, we applied the Mean
Exponential Growth factor of Nearby Orbits \citep[MEGNO, $\Y$][]{Cincotta2003}
which is  a CPU-efficient variant of the Maximal Lyapunov Exponent (MLE)
{useful for analysing stability of planetary systems} with strongly
interacting planets \citep[e.g.][]{Gozdziewski2014}. 

\begin{figure*}
\centerline{
\vbox{
\hbox{
\includegraphics[width=0.33\textwidth]{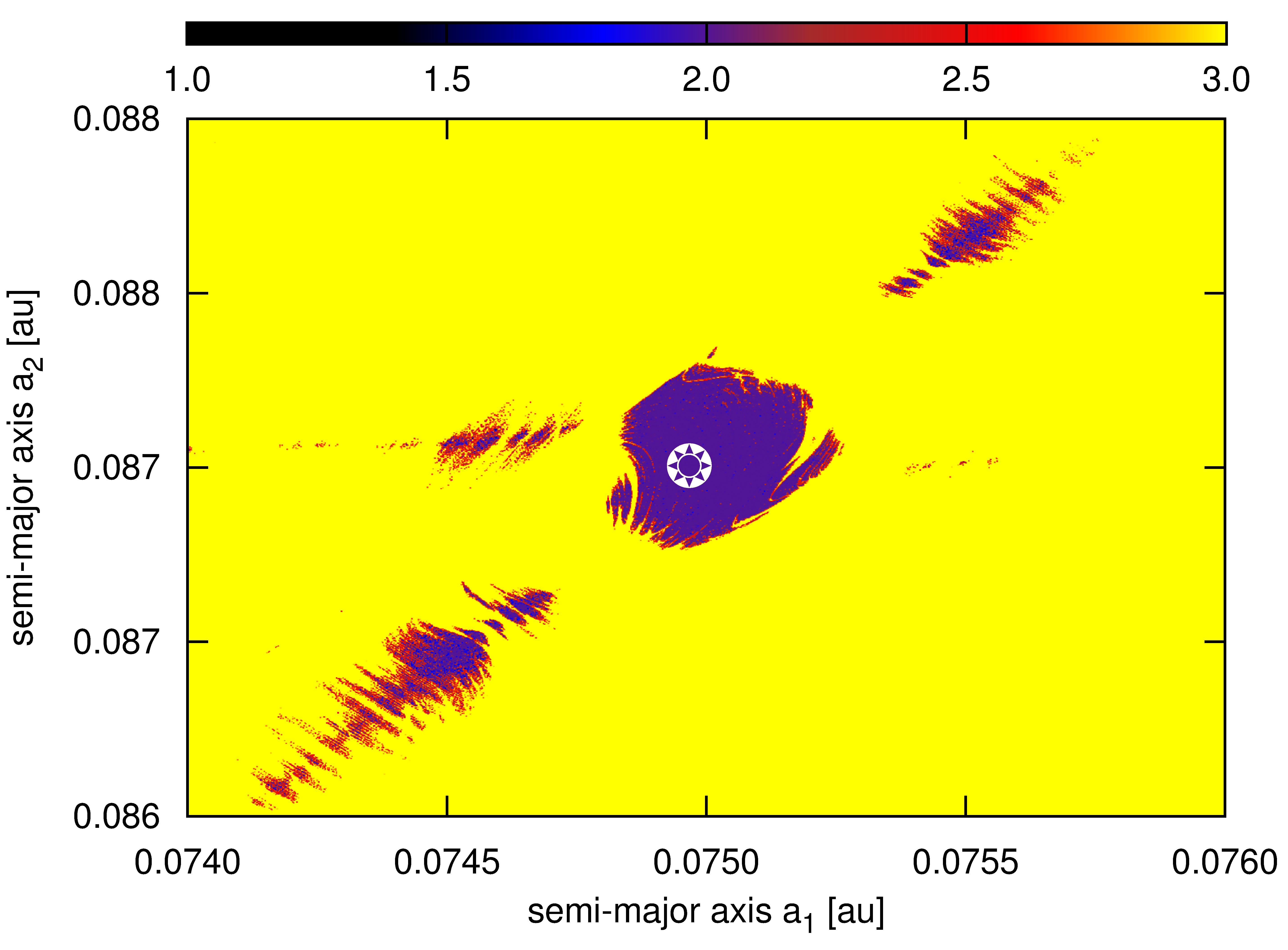}
\includegraphics[width=0.33\textwidth]{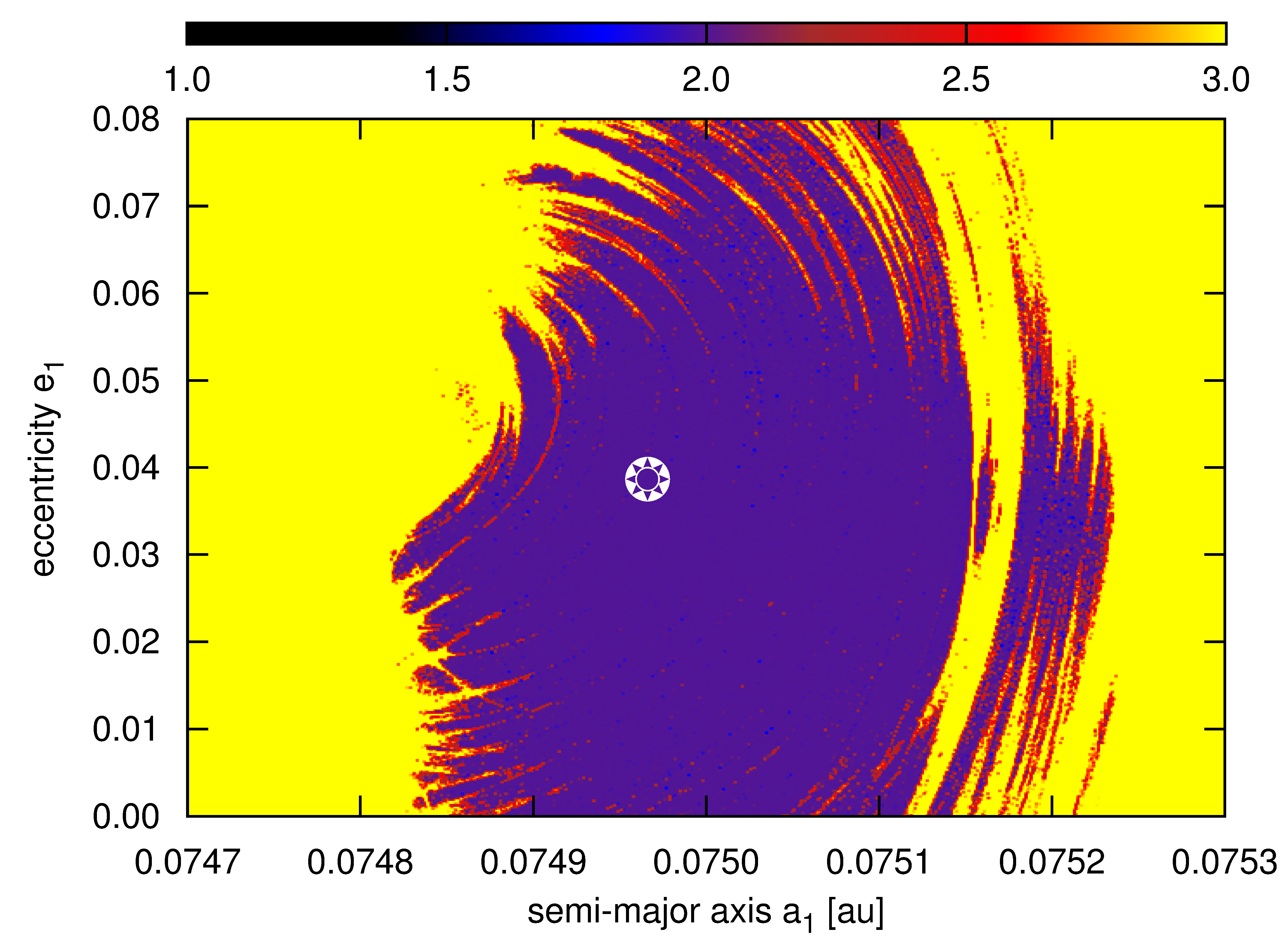}
\includegraphics[width=0.33\textwidth]{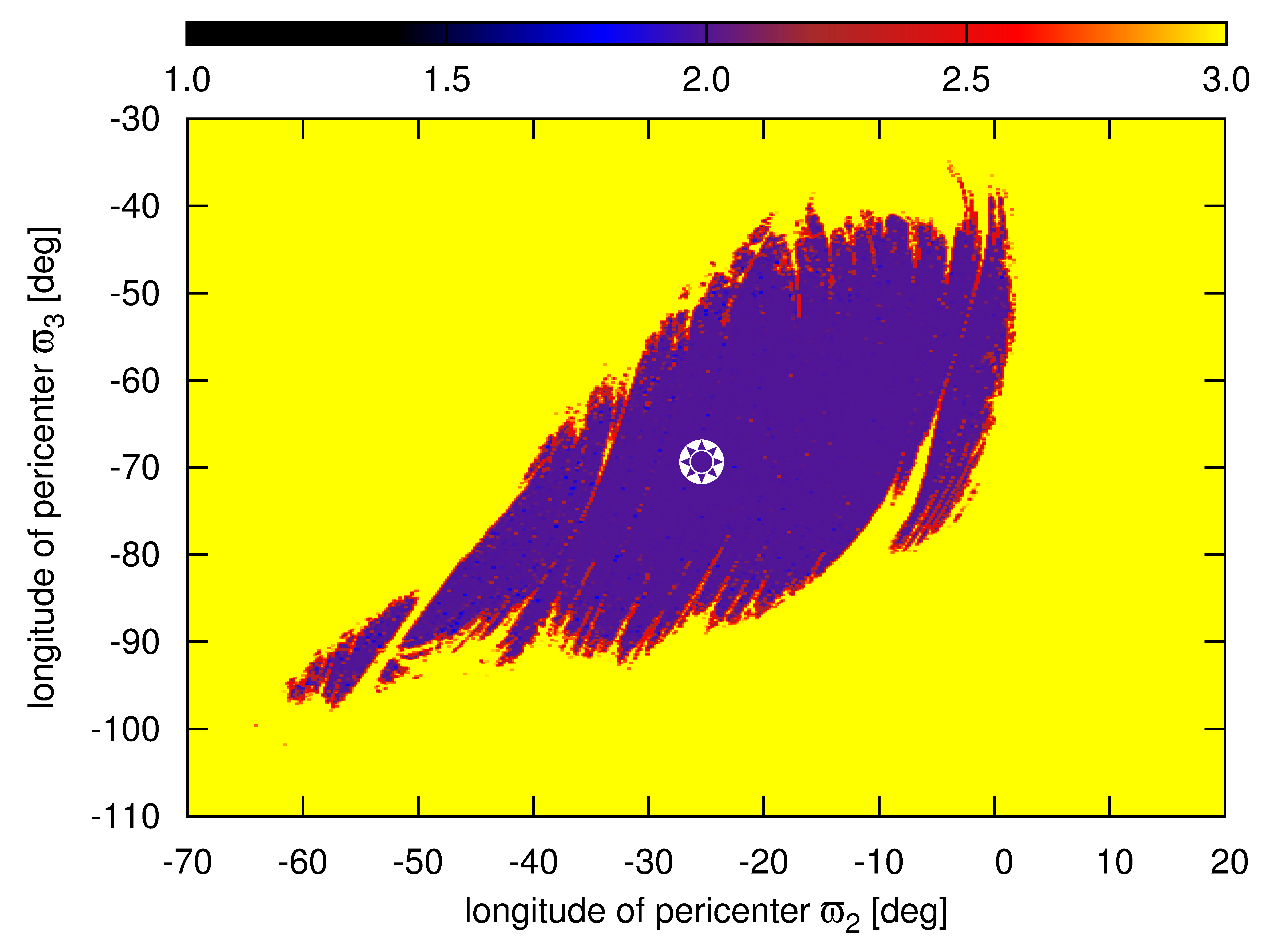}
}
\hbox{
\includegraphics[width=0.33\textwidth]{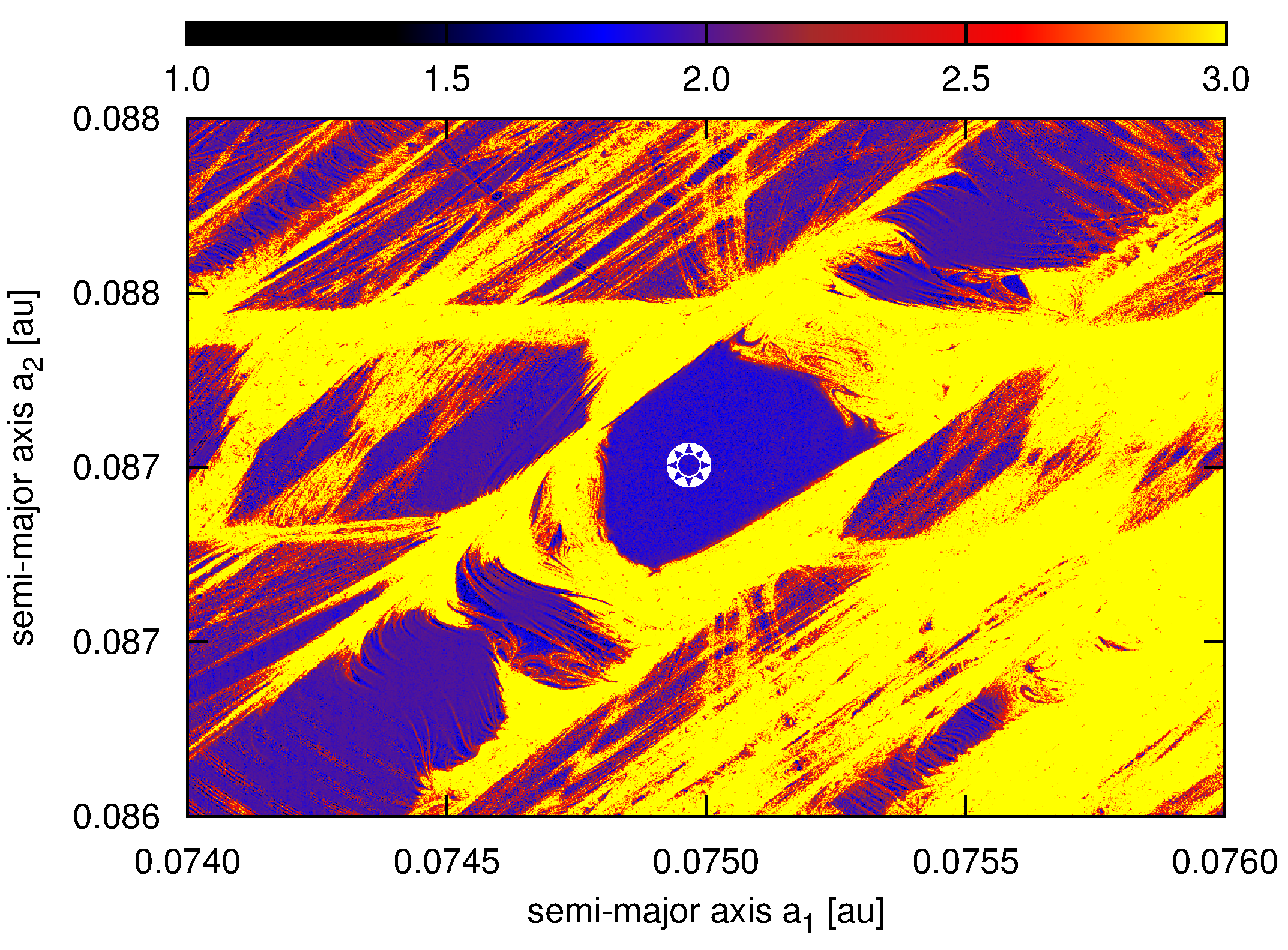}
\includegraphics[width=0.33\textwidth]{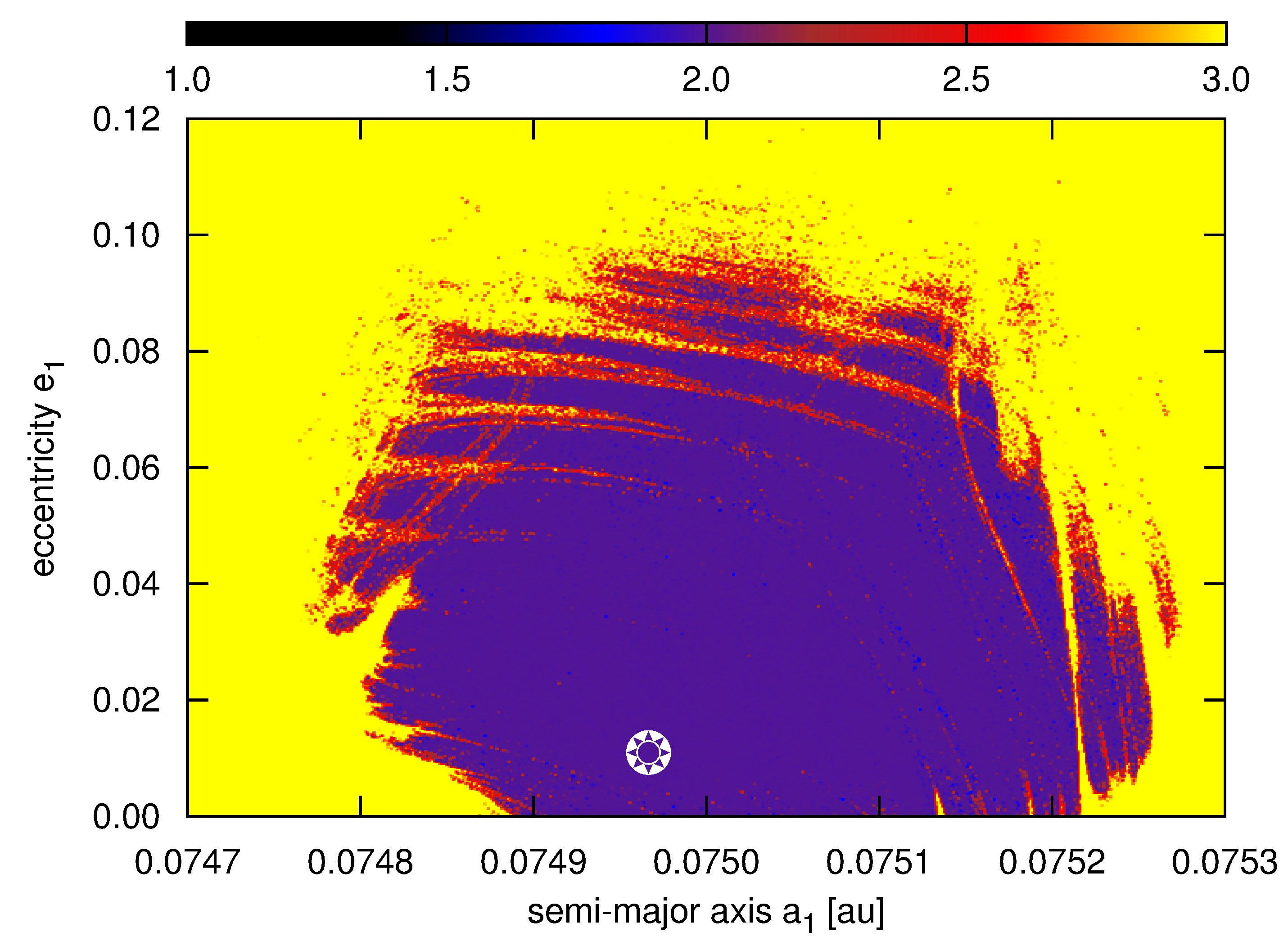}
\includegraphics[width=0.33\textwidth]{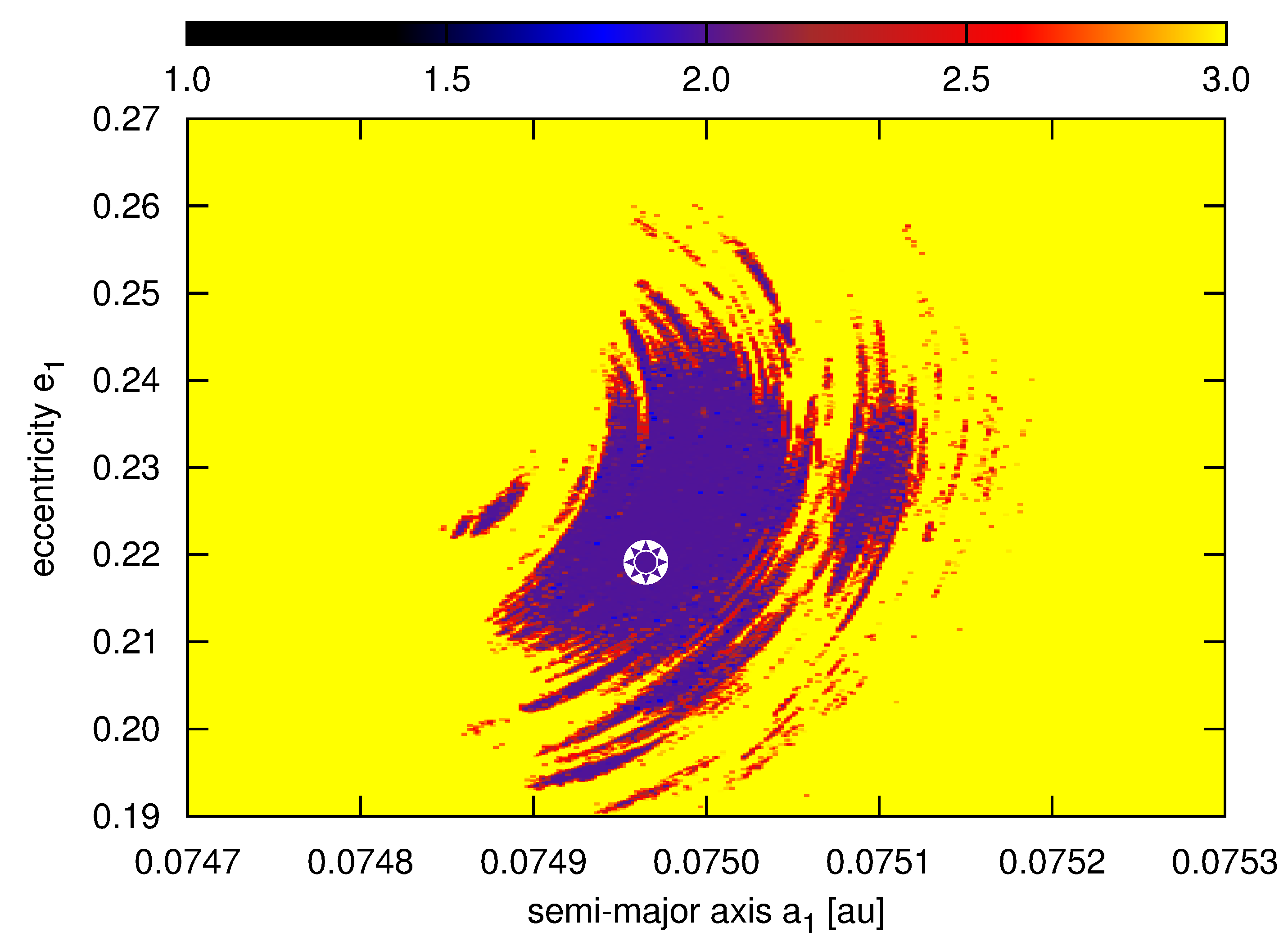}
}
}
}
\caption{
Dynamical maps for the best-fitting three-body MMRs models in
Tab.~\ref{tab:tab1}. The MEGNO $\Y \sim2$ indicates a regular (long-term
stable) solution marked with black/dark blue colour, $\Y$ much larger than $2$,
up to $\gtrsim 5$ indicates a chaotic solution (yellow). The integration time of
each initial condition is 32,000~yr ($\sim10^6 \times \Pd$) besides the
bottom-left panel where the integration time is 80~yr. The asterisk symbol means
the position of the nominal model. {\em The top-row:} dynamical maps for Fit~I
in Tab.~\ref{tab:tab1}. Subsequent maps are for the $(\ab,\ac)$--, $(\ab,\eb)$--
and $(\varpi_{2},\varpi_{3}$)--planes. {The bottom-row:} $(\ab,\ac)$--map for
Fit~I with reduced integration time, $(\ab,\eb)$--plane for Fit~II, and
$(\ab,\eb)$--map for the true Laplace resonance in moderate--eccentricity
region. See the text for details.
}
\label{fig:fig4}
\end{figure*}

We computed 2-dim dynamical maps in the neighborhood of the best-fitting
solutions to show the MMRs structure of the phase-space. The top left-hand panel
of Fig.~\ref{fig:fig4} shows the dynamical map in the $(\ab, \ac)$--plane for
the best-fitting true Laplace MMR (Tab.~\ref{tab:tab1}). (A map for the second,
MMR chain mode looks similar). All other orbital elements are kept at their
best-fitting values. For each initial condition at the grid, the equations of
motion where integrated up to $32,000\,\yr$. This corresponds to $\sim10^6
\times \Pd$, which is sufficient to detect short-term chaotic motions for the
MMRs instability time-scale \citep[e.g.][]{Gozdziewski2014}. 

As the dynamical maps show, the observational uncertainties of $a_1$ and $a_2$
$\simeq 10^{-5}\,\au$ are much smaller than the width of stable islands. For the
true Laplace MMR, we found narrower islands of stable motions even for $e \sim
0.22$ and larger (the bottom-right panel in Fig.~\ref{fig:fig4}) consistent with
the posterior {(online material)}. However, such large eccentricities
would be difficult to explain in the framework of the present state of the
planet formation theory (Sect.~4).
 
We note that stable solutions found for masses $\sim4~\mE$ concur with an
empirical mass-radius relation $m_p/\mE = (R_p/\RE)^{2.06}$ by
\cite{Fabrycky2014}. It provides $\mb \simeq 4.0\,\mE$, $\mc \simeq 4.7\,\mE$
and $\md \simeq 3.2\,\mE$. For second type of the best-fitting models the masses
$\sim10~\mE$ and eccentricities $\sim0.2$ --- we did not find long-term stable
solutions. Yet these models provide $L\sim0.030$~d, which means worse solutions
than derived for the smaller mass range. 

A dynamical character of the system is illustrated in the $(\ab,\ac)$-map 
in the bottom-left panel of Fig.~\ref{fig:fig4} which has been computed for only
80~years. Already for this time-span a strongly chaotic structure appears,
revealing dominant, wide strips of two-body and three-body MMRs that intersect
in a central island of stable configurations that corresponds to the Laplace
MMR. Such a structure may be interpreted as the Arnold web emerging due to MMRs
overlap. A build-up of this structure is illustrated in the $(\ab,\ac)$-maps
(the left column in Fig.~\ref{fig:fig4}). After a sufficient saturation time
($\sim4,000$~yr and longer), only narrow stable regions remain in the top
left-hand map that indicates the Chirikov regime of the chaotic dynamics
\citep{Froeschle2000,Guzzo2006}. In that case a chaotic diffusion may lead to a
random-walk ``wandering'' of solutions along the resonances. This effect leads
to relatively fast and significant changes of the dynamical actions (semi-major
axes). Chaotic models are unstable in the Lagrangian, geometrical sense,
particularly in regions of moderate and large eccentricities. The Arnold web is
one more feature of a strongly resonant system. {Rigorously stable
best-fitting solutions found in this paper are peculiar, since as we confirm
here, recent studies \citep{Wang2012,Showalter2015} indicate that three-body
resonances exhibit very short Lyapunov times that may cause a rapid
instability.}

%_______________________________________________________________________________
%
\section{The true three-body MMR via migration?}
%_______________________________________________________________________________
%
{A formation of the three-body MMRs has been recently studied
by e.g. \cite{Libert2011,Quillen2011,Quillen2014,Batygin2015}}.
It is widely accepted that short-period planets form in protoplanetary discs at
distances wider than the ones observed now and then migrate inwards due to
planet-disc interactions. It is known that convergent migration of a few planets
leads to formation of chains of MMRs \citep[e.g.,][]{Papaloizou2010}. 
From this perspective the true three-body resonance seems to be unexpected. A
lack of two-body MMRs is not the only surprising feature of this configuration.
Also mutual orientations of the apsidal lines (apsides of subsequent orbits are
aligned) differ from a typical outcome of the migration of three-planet systems,
i.e., apsides anti-aligned in a low eccentricity regime \citep{Papaloizou2014}.
Although the orientations may be different if merging and scattering of
initially larger number of planets (or protoplanetary cores) are considered
\citep{Terquem2007}.

Explaining how the true three-body MMR could be formed via migration is well
beyond the scope of this Letter, as many different disc parameters and initial
orbits should be tested. Instead, we tried to verify if the {\em conditio sine
qua non} of such scenario is fulfilled. We tested if chosen representative
configurations are stable against migration, i.e., whether or not after adding
forces mimicking the migration and circularization to the equations of motion
\citep{Moore2013} the system stays in the true three-body MMR or leaves it
(possibly moving towards a chain of two-body MMRs). 
%\hide{
%
%\begin{equation}
%f_{\idm{migr}, i} = -\frac{\vec{v}_i}{2\,\tau_i} - 
%                         \frac{\vec{v}_i - \vec{v}_{c,i}}{K\,\tau_i},
%\end{equation}
%
%where $\vec{v}$ is the velocity of $i$-th planet, $\vec{v}_{c,i}$ is the
%Keplerian velocity of a~planet in a circular orbit with radius $r_i$ (an
%astrocentric distance of $i$-th planet), $\tau_i$ is the time-scale of migration
%of the $i$-th planet, $K$ is the ratio of the migration and circularization
%time-scales.}

We checked that regardless of whether the migration is convergent or divergent
the system leaves the resonance. The parameter $K$ which is the ratio of the
migration and circularization time-scales was being changed in a range of $[0.1,
100]$. For $K \gtrsim 1$ the systems tend towards a chain of two-body MMRs with
anti-aligned apsides of subsequent orbits. Naturally, for the divergent
migration the period ratios increase and the systems leave the chain. For $K
\lesssim 1$ the systems self-disrupt. Those tests suggest that the true
three-body MMR will be a challenge for the resonances formation theory. 

%_______________________________________________________________________________
%
\section{Conclusions}
%_______________________________________________________________________________
%
The orbital periods of the Kepler-60 planetary system exhibit close
commensurabilities which may be interpreted as a chain of two-body, first-order
MMRs or the true three-body MMR with none of the two-body MMRs critical angles
librating. We found a strong observational indication that the zero-th order
three-body MMR $1:-2:1$ is present. Regardless of its type, it could be
considered as the generalized Laplace resonance \citep{Papaloizou2014}. It is
characterized by $\sim10^{\circ}$--amplitude libration around $\sim45^{\circ}$
libration center.

The TTV series imply a very complex structure of the phase space of the system.
Strong mutual perturbations between $\sim4\,\mE$ super-Earths lead to the
Chirikov regime of the dynamics which is governed by chaotic diffusion due to
the overlap of two- and three-body MMRs. Long-term stable orbital configurations
are confined to isolated islands associated with the MMRs. 
A past putative migration probably rules out the true three-body MMR
since the resonance is not robust against the migration. The most likely state
of Kepler-60 system is a chain of two-body MMRs with all critical angles
librating with small amplitudes. The lightcurve has low signal-to-noise ratio,
similar with many other multiple-systems in the \kepler{} sample. Therefore
constraining a particular type of the Laplace MMR in the Kepler-60 system
{with the present data is not likely}. It remains an open problem whose
solution may shed more light on the formation of this intriguing system.

%_____________________________________________________________________________
%
\section{Acknowledgements}
%_______________________________________________________________________________
%
We thank the anonymous reviewer for comments that improved this paper.
This work has been supported by Polish National Science Centre MAESTRO grant
DEC-2012/06/A/ST9/00276. K.G. thanks the Pozna\'n Supercomputer and Network
Centre (PCSS, Poland) for computing resources (grant No. 195) and
a generous support.
%_______________________________________________________________________________
%
\bibliographystyle{mn2e}
\bibliography{ms}

\begin{thebibliography}{}

\bibitem[\protect\citeauthoryear{{Agol}, {Steffen}, {Sari} \&
  {Clarkson}}{{Agol} et~al.}{2005}]{Agol2005}
{Agol} E.,  {Steffen} J.,  {Sari} R.,    {Clarkson} W.,  2005, \mnras, 359, 567

\bibitem[\protect\citeauthoryear{{Baluev}}{{Baluev}}{2009}]{Baluev2009}
{Baluev} R.~V.,  2009, \mnras, 393, 969

\bibitem[\protect\citeauthoryear{{Batygin}, {Deck} \& {Holman}}{{Batygin}
  et~al.}{2015}]{Batygin2015}
{Batygin} K.,  {Deck} K.~M.,    {Holman} M.~J.,  2015, \aj, 149, 167

\bibitem[\protect\citeauthoryear{{Carter}, {Fabrycky}, {Ragozzine}, {Holman},
  {Quinn} \& et al.}{{Carter} et~al.}{2011}]{Carter2011}
{Carter} J.~A.,  {Fabrycky} D.~C.,  {Ragozzine} D.,  {Holman} M.~J.,  {Quinn}
  S.~N.,    et al. 2011, Science, 331, 562

\bibitem[\protect\citeauthoryear{{Charbonneau}}{{Charbonneau}}{1995}]{Charbonneau1995}
{Charbonneau} P.,  1995, \apjs, 101, 309

\bibitem[\protect\citeauthoryear{{Cincotta}, {Giordano} \&
  {Sim{\'o}}}{{Cincotta} et~al.}{2003}]{Cincotta2003}
{Cincotta} P.~M.,  {Giordano} C.~M.,    {Sim{\'o}} C.,  2003, Physica D
  Nonlinear Phenomena, 182, 151

\bibitem[\protect\citeauthoryear{{Deck}, {Agol}, {Holman} \&
  {Nesvorn{\'y}}}{{Deck} et~al.}{2014}]{Deck2014}
{Deck} K.~M.,  {Agol} E.,  {Holman} M.~J.,    {Nesvorn{\'y}} D.,  2014, \apj,
  787, 132

\bibitem[\protect\citeauthoryear{{Fabrycky}, {Lissauer}, {Ragozzine}, {Rowe},
  {Steffen}, {Agol}, {Barclay} \& et al.}{{Fabrycky}
  et~al.}{2014}]{Fabrycky2014}
{Fabrycky} D.~C.,  {Lissauer} J.~J.,  {Ragozzine} D.,  {Rowe} J.~F.,  {Steffen}
  J.~H.,  {Agol} E.,  {Barclay} T.,    et al. 2014, \apj, 790, 146

\bibitem[\protect\citeauthoryear{{Foreman-Mackey}, {Hogg}, {Lang} \&
  {Goodman}}{{Foreman-Mackey} et~al.}{2013}]{Foreman2014}
{Foreman-Mackey} D.,  {Hogg} D.~W.,  {Lang} D.,    {Goodman} J.,  2013, \pasp,
  125, 306

\bibitem[\protect\citeauthoryear{{Froeschl{\'e}}, {Guzzo} \&
  {Lega}}{{Froeschl{\'e}} et~al.}{2000}]{Froeschle2000}
{Froeschl{\'e}} C.,  {Guzzo} M.,    {Lega} E.,  2000, Science, 289, 2108

\bibitem[\protect\citeauthoryear{{Goodman} \& {Weare}}{{Goodman} \&
  {Weare}}{2010}]{Goodman2010}
{Goodman} J.,  {Weare} J.,  2010, Comm. Apl. Math and Comp. Sci., 1, 65

\bibitem[\protect\citeauthoryear{{Go{\'z}dziewski} \&
  {Migaszewski}}{{Go{\'z}dziewski} \& {Migaszewski}}{2014}]{Gozdziewski2014}
{Go{\'z}dziewski} K.,  {Migaszewski} C.,  2014, \mnras, 440, 3140

\bibitem[\protect\citeauthoryear{{Guzzo}}{{Guzzo}}{2006}]{Guzzo2006}
{Guzzo} M.,  2006, Icarus, 181, 475

\bibitem[\protect\citeauthoryear{{Lee}, {Fabrycky} \& {Lin}}{{Lee}
  et~al.}{2013}]{Fabrycky2013}
{Lee} M.~H.,  {Fabrycky} D.,    {Lin} D.~N.~C.,  2013, \apj, 774, 52

\bibitem[\protect\citeauthoryear{{Libert} \& {Tsiganis}}{{Libert} \&
  {Tsiganis}}{2011}]{Libert2011}
{Libert} A.-S.,  {Tsiganis} K.,  2011, Celestial Mechanics and Dynamical
  Astronomy, 111, 201

\bibitem[\protect\citeauthoryear{{Marcy}, {Butler}, {Fischer}, {Vogt},
  {Lissauer} \& {Rivera}}{{Marcy} et~al.}{2001}]{Marcy2001}
{Marcy} G.~W.,  {Butler} R.~P.,  {Fischer} D.,  {Vogt} S.~S.,  {Lissauer}
  J.~J.,    {Rivera} E.~J.,  2001, \apj, 556, 296

\bibitem[\protect\citeauthoryear{{Marois}, {Zuckerman}, {Konopacky},
  {Macintosh} \& {Barman}}{{Marois} et~al.}{2010}]{Marois2010}
{Marois} C.,  {Zuckerman} B.,  {Konopacky} Q.~M.,  {Macintosh} B.,    {Barman}
  T.,  2010, \nat, 468, 1080

\bibitem[\protect\citeauthoryear{{Mart{\'{\i}}}, {Giuppone} \&
  {Beaug{\'e}}}{{Mart{\'{\i}}} et~al.}{2013}]{Marti2013}
{Mart{\'{\i}}} J.~G.,  {Giuppone} C.~A.,    {Beaug{\'e}} C.,  2013, \mnras,
  433, 928

\bibitem[\protect\citeauthoryear{{Moore}, {Hasan} \& {Quillen}}{{Moore}
  et~al.}{2013}]{Moore2013}
{Moore} A.,  {Hasan} I.,    {Quillen} A.~C.,  2013, \mnras, 432, 1196

\bibitem[\protect\citeauthoryear{{Mullally et al.}}{{Mullally et
  al.}}{2015}]{Mullally2015}
{Mullally et al.} 2015, \apjs, 217, 31

\bibitem[\protect\citeauthoryear{{Murray} \& {Holman}}{{Murray} \&
  {Holman}}{1999}]{Murray1999}
{Murray} N.,  {Holman} M.,  1999, Science, 283, 1877

\bibitem[\protect\citeauthoryear{{Papaloizou}}{{Papaloizou}}{2015}]{Papaloizou2014}
{Papaloizou} J.~C.~B.,  2015, International Journal of Astrobiology, 14, 291

\bibitem[\protect\citeauthoryear{{Papaloizou} \& {Terquem}}{{Papaloizou} \&
  {Terquem}}{2010}]{Papaloizou2010}
{Papaloizou} J.~C.~B.,  {Terquem} C.,  2010, \mnras, 405, 573

\bibitem[\protect\citeauthoryear{{Quillen}}{{Quillen}}{2011}]{Quillen2011}
{Quillen} A.~C.,  2011, \mnras, 418, 1043

\bibitem[\protect\citeauthoryear{{Quillen} \& {French}}{{Quillen} \&
  {French}}{2014}]{Quillen2014}
{Quillen} A.~C.,  {French} R.~S.,  2014, \mnras, 445, 3959

\bibitem[\protect\citeauthoryear{{Rivera}, {Laughlin}, {Butler}, {Vogt},
  {Haghighipour} \& {Meschiari}}{{Rivera} et~al.}{2010}]{Rivera2010}
{Rivera} E.~J.,  {Laughlin} G.,  {Butler} R.~P.,  {Vogt} S.~S.,  {Haghighipour}
  N.,    {Meschiari} S.,  2010, \apj, 719, 890

\bibitem[\protect\citeauthoryear{{Rowe}, {Coughlin}, {Antoci}, {Barclay},
  {Batalha}, {Borucki}, {Burke} \& et al.}{{Rowe} et~al.}{2015}]{Rowe2015}
{Rowe} J.~F.,  {Coughlin} J.~L.,  {Antoci} V.,  {Barclay} T.,  {Batalha} N.~M.,
   {Borucki} W.~J.,  {Burke} C.~J.,    et al. 2015, \apjs, 217, 16

\bibitem[\protect\citeauthoryear{{Ruci{\'n}ski}, {Izzo} \&
  {Biscani}}{{Ruci{\'n}ski} et~al.}{2010}]{Izzo2010}
{Ruci{\'n}ski} M.,  {Izzo} D.,    {Biscani} F.,  2010, Parallel Computing, 10,
  555

\bibitem[\protect\citeauthoryear{{Showalter} \& {Hamilton}}{{Showalter} \&
  {Hamilton}}{2015}]{Showalter2015}
{Showalter} M.~R.,  {Hamilton} D.~P.,  2015, \nat, 522, 45

\bibitem[\protect\citeauthoryear{{Sinclair}}{{Sinclair}}{1975}]{Sinclair1975}
{Sinclair} A.~T.,  1975, \mnras, 171, 59

\bibitem[\protect\citeauthoryear{{Smirnov} \& {Shevchenko}}{{Smirnov} \&
  {Shevchenko}}{2013}]{Smirnov2013}
{Smirnov} E.~A.,  {Shevchenko} I.~I.,  2013, \icarus, 222, 220

\bibitem[\protect\citeauthoryear{{Steffen}, {Fabrycky}, {Ford}, {Carter},
  {D{\'e}sert} \& et al.}{{Steffen} et~al.}{2012}]{Steffen2012}
{Steffen} J.~H.,  {Fabrycky} D.~C.,  {Ford} E.~B.,  {Carter} J.~A.,
  {D{\'e}sert} J.-M.,    et al. 2012, \mnras, 421, 2342

\bibitem[\protect\citeauthoryear{{Terquem} \& {Papaloizou}}{{Terquem} \&
  {Papaloizou}}{2007}]{Terquem2007}
{Terquem} C.,  {Papaloizou} J.~C.~B.,  2007, \apj, 654, 1110

\bibitem[\protect\citeauthoryear{{Wang}, {Ji} \& {Zhou}}{{Wang}
  et~al.}{2012}]{Wang2012}
{Wang} S.,  {Ji} J.,    {Zhou} J.-L.,  2012, \apj, 753, 170

\end{thebibliography}
\label{lastpage}
%_______________________________________________________________________________
%
\section*{\large \bfseries Supplementary online material}
%_______________________________________________________________________________
%
\vspace*{-2in}
\begin{figure*}
\centerline{\includegraphics[width=1.0\textwidth]{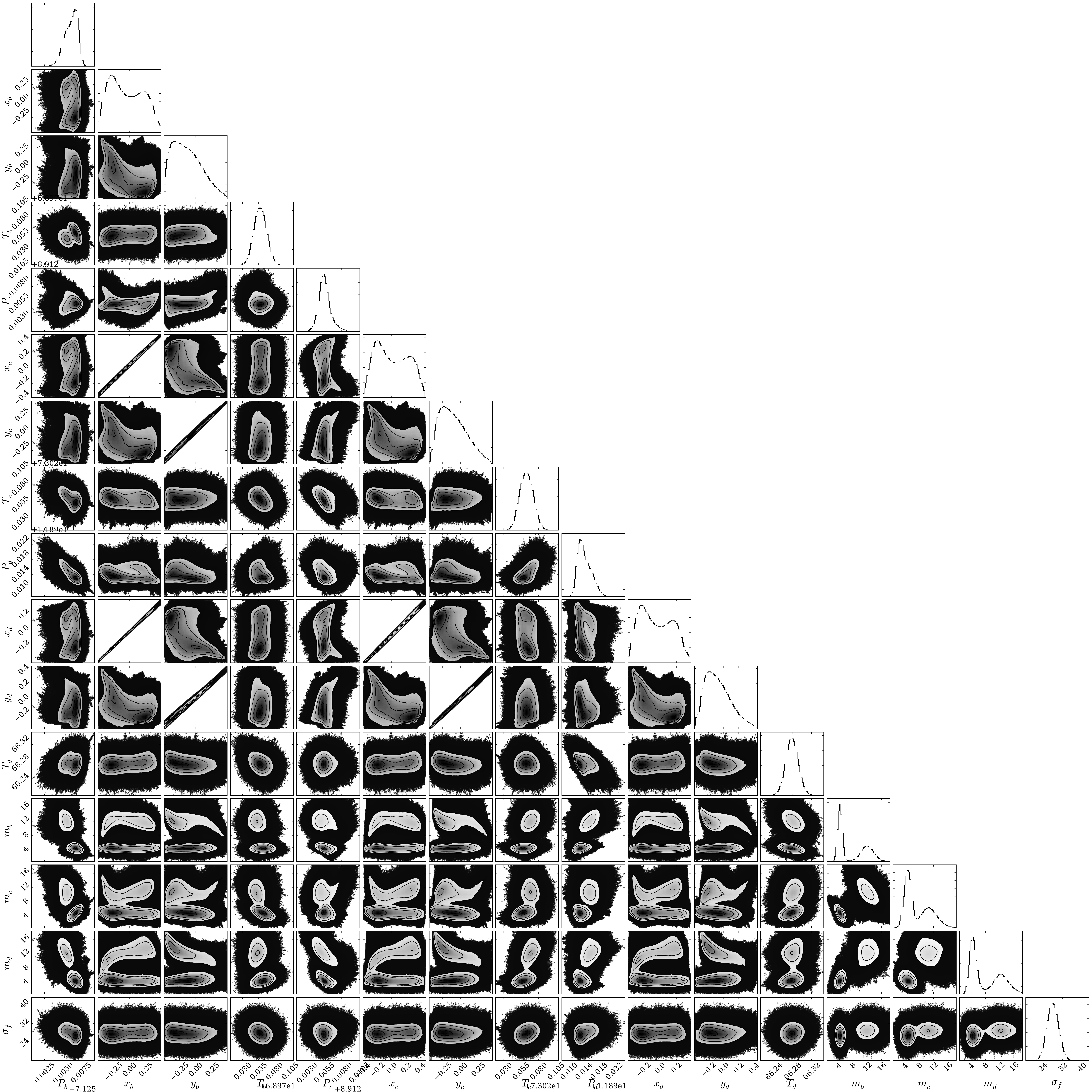}}
\caption{
One-- and two--dimensional projections of the posterior probability distribution
for all free parameters of the TTV model. The MCMC chain length {is 1,024,000
iterations in each of 512 different instances selected in a small ball around a
solution found with the genetic algorithms}. First transit epochs $T_{b,c,d}$ and
orbital periods $P_{b,c,d}$ are expressed in days, planetary masses $m_{b,c,d}$
are expressed in Earth masses, and the uncertainty correction term $\sigma_f$ is
given in minutes ($b\equiv 1,c\equiv 2,d\equiv 3$ are indices of subsequent
planets).
}
\label{fig:fig5}
\end{figure*}

\end{document}